\documentclass{article}

\PassOptionsToPackage{numbers,sort&compress}{natbib}
\usepackage[creativeai, final]{neurips_2025}

\usepackage[utf8]{inputenc} 
\usepackage[T1]{fontenc}    
\usepackage{hyperref}       
\usepackage{url}            
\usepackage{booktabs}       
\usepackage{amsfonts}       
\usepackage{amsmath}        
\usepackage{amssymb}        
\usepackage{nicefrac}       
\usepackage{microtype}      
\usepackage{xcolor}         
\usepackage{graphicx}       
\usepackage{enumitem}       
\usepackage{tabularx}
\usepackage{threeparttable}
\usepackage{siunitx}
\title{BioArtlas: Computational Clustering of Multi-Dimensional Complexity in Bioart}

\author{
  Joonhyung Bae \\
  Graduate School of Culture Technology \\
  KAIST (Korea Advanced Institute of Science and Technology) \\
  Daejeon, South Korea \\
  \texttt{jh.bae@kaist.ac.kr}
}

\begin{document}
\maketitle

\begin{abstract}
Bioart brings living material into artistic practice, where a single work can be at once an aesthetic object, a scientific instrument, and an ethical provocation. Traditional categories sort such works along one axis at a time, which flattens the very hybridity that defines the field and leaves curators no way to compare works across many dimensions together. I introduce BioArtlas, a computational atlas that represents each bioartwork along many curated dimensions at once and organizes the field by conceptual similarity rather than by medium or chronology. My method embeds the keywords of all 81 works on each of thirteen interpretive axes, groups related concepts into a shared codebook that tames inconsistent terminology, and then searches systematically for a clustering that is both statistically clean and interpretable. Among the methods that place every work on the map, agglomerative clustering separates the field far more cleanly than the usual k-means baseline (silhouette 0.664 versus 0.483), whereas density-based methods reach higher scores only by discarding most of the corpus as noise. By separating rigorous analysis from public storytelling, BioArtlas turns the tangled complexity of bioart into a navigable landscape, openly available as an interactive interface (\url{https://www.bioartlas.com}) and dataset (\url{https://github.com/joonhyungbae/BioArtlas}).
\end{abstract}

\section{Introduction}
Bioart brings living material into the studio. Artists in this field cultivate cells and tissues, engineer organisms, and engage living systems to ask about identity, ethics, embodiment, and the line between the natural and the made \cite{kac1998transgenic,kac2005telepresence,catts2008ethics,stelarc2005monograph}. Over four decades it has grown into an international field with its own awards and institutions \cite{yetisen2015bioart,anker2004molecular,mitchell2010bioart}.

Scholars and curators have largely understood this field one lens at a time, organizing works by their medium, by the biological techniques they employ, or by the ethical stance they take toward living systems \cite{hauser2005,zkm-biomedia,catts2008ethics}. Each lens is illuminating on its own.

Yet a single bioartwork is rarely one thing. The same piece can be an aesthetic object, a laboratory experiment, an ethical provocation, and a political statement at once, and any single-axis scheme forces it into a box that hides most of what it is \cite{latour-reassembling,haraway2008species}. A medium-first reading mutes the ethics, an ethics-first reading mutes the aesthetics, and neither lets us see the field as a whole. What is missing is a way to lay bioart out and ask which works are conceptually close across all these dimensions at once.

BioArtlas starts from the opposite end. Rather than choosing one axis, I describe each work along many interpretive dimensions and let the works reveal which ones belong together. The intuition is simple. If two pieces share concerns across their materials, methods, ethics, and aesthetics, they should sit near each other even when they use different media or were made decades apart. To do this, I turn each artwork into an axis-aware representation, reconcile inconsistent vocabulary through a shared concept codebook, and search broadly for a clustering that is both clean and readable, building on cultural analytics and sentence embeddings \cite{jones2025toward,manovich2009cultural,reimers2019sentence}.

This work makes four contributions.
\begin{itemize}[topsep=2pt,itemsep=1pt,parsep=0pt,leftmargin=1.3em]
\item I introduce an axis-aware representation that compares works across many interpretive dimensions while keeping each dimension distinct.
\item I build a domain-informed concept codebook so that clustering reflects shared ideas rather than accidental wording.
\item I treat model selection as a systematic search, and separate the configuration chosen for analytical rigor from the simpler labeling used for public communication.
\item I surface four recurring ways the field organizes itself, from an artist's methodological signature to conceptual kinships that span decades.
\end{itemize}

\section{Dataset}
The corpus was built to select target artists and representative works by synthesizing \emph{award-institution-platform} indicators conferring international visibility from major bioart awards and exhibitions, including \emph{Prix Ars Electronica}, \emph{Bio Art \& Design Award}, ZKM's \emph{BioMedia}, MIT List's \emph{Symbionts}, STARTS Prize, and ISEA archives \cite{ars-artificial-life,bad-award,zkm-biomedia,mitlist-symbionts,starts-prize,isea-archives}. Artist and artwork selection used multi-layered criteria, namely art-historical significance (e.g., Davis, Kac), technological innovation, field contribution, and conceptual clarity \cite{kac1998transgenic,kac2005telepresence}.

The \textbf{13 analytic axes} (Table~\ref{tab:axis-keywords}) were derived via literature review \cite{kac1998transgenic,catts2008ethics,hauser2005,zkm-biomedia,latour-reassembling,haraway2008species} and inductive corpus analysis. Coverage and non-redundancy were emphasized during axis development. Observed pairwise correlations were low-to-moderate, consistent with complementary dimensions. All source materials are public, non-personal, and attributed. Only metadata and derived artifacts are released. Policies follow FAIR/CARE guidelines. Takedown requests are honored. Axis definitions and labeling protocols were predetermined using a codebook with comprehensive annotation criteria, shaped by the author's dual role as a professional artist and an AI researcher.  Although the existing annotations are conducted by a single annotator, the systematic codebook facilitates future validation by many annotators and the evaluation of inter-rater reliability.

 This study provides a systematically annotated bioart dataset, featuring systematic evaluations across 13 analytical parameters.  Although constrained in scope, it provides a preliminary basis for computational methodologies in bioart analysis and associated cultural fields.  Table~\ref{tab:dataset-stats} offers a detailed summary of the dataset's composition, illustrating the significant scale and diversity of the corpus, which contains 770 distinct keywords across 81 artworks.

\begin{table}[!htbp]
\vspace{-5pt}
\centering
\caption{Dataset Summary}
\label{tab:dataset-stats}
\begingroup
\setlength{\tabcolsep}{4pt}\renewcommand{\arraystretch}{0.95}\scriptsize
\begin{tabularx}{\linewidth}{>{\raggedright\arraybackslash}X
                        >{\raggedright\arraybackslash}X}
\toprule
\textbf{Dataset} & \textbf{Keywords} \\
\midrule
\begin{tabular}{@{}ll@{}}
Total works & 81 \\
Total artists & 33 \\
Temporal coverage & 1976--2022 \\
\end{tabular}
&
\begin{tabular}{@{}ll@{}}
Unique keywords & 770 \\
Mean / work & 28.2 \\
Assignments & 2285 \\
Analytic axes & 13 \\
\end{tabular}
\\
\bottomrule
\end{tabularx}
\endgroup
\vspace{-5pt}
\end{table}

\begin{table}[!htbp]
\centering
\caption{Keyword statistics for the 13 analytic axes, with the top three keywords per axis.}
\label{tab:axis-keywords}
\begingroup
\setlength{\tabcolsep}{4pt}
\renewcommand{\arraystretch}{0.95}
\scriptsize
\begin{tabularx}{\linewidth}{l c c p{6cm}}
\toprule
Axis & Unique & Average & Top (up to three) \\
\midrule
Materiality & 98 & 2.58 & Plant, Composite materials, Data \\
Methodology & 88 & 2.40 & Cell culture, Data visualization, Biosensing \\
Actor Relations \& Configurations & 62 & 2.17 & Artist-led, Autonomous bio processes, Interspecies co-creation \\
Ethical Approach & 53 & 2.16 & Reflective, Relational ethics, Symbiotic \\
Aesthetic Strategy & 76 & 2.57 & Conceptual, Uncanny, Biological morphology \\
Epistemic Function & 54 & 2.21 & Social criticism, Knowledge production, Future proposal \\
Philosophical Stance & 47 & 2.23 & New materialism, Posthumanism, Relational ontology \\
Social Context & 44 & 2.17 & Gallery, Laboratory \\
Audience Engagement & 53 & 2.00 & Observational, Interpretive engagement, Contemplative \\
Temporal Scale & 44 & 1.53 & Continuous, Short-term exhibition, Evolutionary \\
Spatial Scale & 56 & 1.48 & Installation, Human body size, Individual unit \\
Power and Capital Critique & 61 & 1.79 & Institutional criticism, Biopolitics, Biocapital \\
Documentation \& Representation & 85 & 2.91 & Photographic records, Material residue, Data viz \\
\bottomrule
\end{tabularx}
\endgroup
\end{table}

The dataset includes 33 artists and collectives (Table~\ref{tab:artists-list}), balancing diversity with focused analysis of key practitioners.

\begin{table}[!htbp]
\vspace{-10pt}
\centering
\caption{List of 33 artists and collectives in the dataset, with number of works shown in parentheses.}
\label{tab:artists-list}
\begingroup
\setlength{\tabcolsep}{5pt}\renewcommand{\arraystretch}{0.95}\scriptsize
\begin{tabular}{p{12.8cm}}
\toprule
\textbf{Artist Names (Number of Works)} \\
\midrule
Joe Davis (4), Stelarc (4), George Gessert (2), Eduardo Kac (3), Oron Catts \& Ionat Zurr (3), Wim Delvoye (2), Art Orienté Objet (1), HeHe (1), Zbigniew Oksiuta (1), Paul Vanouse (3), Anna Dumitriu (2), Center for Genomic Gastronomy (2), Charlotte Jarvis (2), Heather Dewey-Hagborg (2), Jalila Essaïdi (2), Marta de Menezes (2), Špela Petrič (2), Agi Haines (1), Maja Smrekar (1), Ani Liu (5), Alexandra Daisy Ginsberg (4), Anicka Yi (4), Candice Lin (3), Claire Pentecost (3), Dasha Tsapenko (3), Jenna Sutela (3), Jes Fan (3), Cecilia Jonsson (2), Gilberto Esparza (2), Pamela Rosenkranz (2), Xandra van der Eijk (2), Amy Karle (1), Michael Sedbon (1) \\
\bottomrule
\end{tabular}
\endgroup
\vspace{-12pt}
\end{table}

\section{Method}

\subsection{Two-Stage Representation Process}
The multidimensional nature of bioart, encompassing aesthetic, ethical, material, and intellectual aspects, necessitates transcending singular embedding techniques that may obscure diverse semantic components. My solution tackles these problems via a methodical two-stage process that maintains axis-specific semantics while facilitating cross-dimensional comparison. The method establishes durable semantic anchors by constructing a codebook that organizes related concepts into cohesive clusters, maintaining consistency among artworks.

\textbf{Stage 1: Per-Axis Embedding Aggregation.} For each artwork's 13 axes, assigned keywords are embedded using dual large language models (BGE-large-en-v1.5 \cite{xiao2024cpack} and GTE-large-en-v1.5 \cite{zhang2024mgte,li2023towards}), both strong performers on standard text-embedding benchmarks \cite{muennighoff2022mteb}, with concatenated representations yielding 2048-dimensional vectors. Keywords within each axis are then averaged to create a single axis-level embedding. For example, if an artwork's \textit{Materiality} axis contains keywords \{\textit{plant}, \textit{data}, \textit{composite-materials}\}, the three 2048-dimensional embeddings are averaged into one 2048-dimensional vector representing that axis. Each artwork thus yields 13 axis-specific embeddings of 2048 dimensions each.

\textbf{Stage 2: Word Codebook and Axis Features.} All unique keywords across the entire dataset are clustered into a \emph{word codebook} using K-means. Prior to clustering, token embeddings undergo PCA (retaining $\geq$95\% cumulative variance) with whitening to stabilize K-means. $K_c$ is \emph{automatically selected} by scanning candidate values and maximizing a silhouette-based objective regularized by penalties for empty/singleton/imbalanced clusters. Each artwork is then represented via per-axis codebook activations, namely counts/one-hot and their TF-IDF (BM25 optional). I also compute \emph{quantized per-axis embeddings} as count-weighted averages of codebook centroids.

The codebook approach addresses polysemy by grouping semantically related terms (e.g., \{\textit{biofabrication}, \textit{tissue-engineering}, \textit{living-materials}\}) into unified concept clusters, enabling more robust similarity computation than raw keyword matching.

\textbf{Final Representation.} I generate TF-IDF weighted cluster counts (L2 normalized), quantized embeddings, binary indicators, and SVD variants. TF-IDF weighted counts achieved superior performance, balancing interpretability with clustering quality.

\subsection{Systematic Sweep Configuration}\label{subsec:conv}
I evaluate multiple representation types across systematic algorithm-space combinations to identify optimal clustering configurations. My approach explores 8 feature representations (TF-IDF variants, quantized embeddings, SVD-reduced features) across 31+ projection spaces (RAW, SVD dimensions 50/100/150, UMAP 4D/8D/16D with 3×3 hyperparameter grids) using 4 clustering algorithms (K-means, Agglomerative, DBSCAN, OPTICS). With $K \in [2,15]$ for partitional methods, this yields 800+ evaluated configurations within computational constraints. Unlike multi-view and multi-kernel fusion or consensus-clustering approaches \cite{yu2025review,zhou2022multi,chen2024multiple,strehl2002cluster,senbabaoglu2014critical}, I do not learn a single fused space but sweep representation-space-algorithm combinations and select for interpretability under atlas constraints.

The building of a cultural atlas necessitates a balance between statistical coherence and interpretive utility, highlighting a fundamental contradiction between optimization objectives.  This challenge is especially pronounced in identifying optimal cluster numbers. Although the gap statistic and silhouette analysis \cite{tibshirani2001estimating} can in principle admit larger partitions (we estimate $K \le 27$ for our corpus of $N=81$), domain-specific constraints in cultural categorization favor cognitively manageable partitions. The range $K \in [2,15]$ reflects this compromise, informed by cluster validation literature \cite{halkidi2001clustering,milligan1985examination,kaufman1990finding}, where excessive granularity obstructs interpretive understanding.   Conventional clustering optimization prioritizes internal cohesion metrics (Silhouette, within-cluster sum of squares), while atlas applications necessitate thorough categorization, cognitive accessibility, and interpretive clarity, often conflicting with statistical optimization.

 This tension is evident in three dimensions. First, \textbf{Completeness vs. Purity}, where density-based methods attain superior silhouette scores by categorizing boundary cases as noise, whereas cultural atlases necessitate extensive landscape representation. Second, \textbf{Statistical Precision vs. Cognitive Load}, where hierarchical methods enhance separation via meticulous partitioning that surpasses human categorical processing capabilities. Third, \textbf{Algorithmic Sophistication vs. Interpretive Transparency}, where sophisticated techniques may identify statistical patterns while concealing semantic distinctions.

Multi-Metric Evaluation Framework.  I resolve these conflicts by doing a thorough assessment utilizing many validation criteria instead of depending on singular measures.  Evaluation incorporates the silhouette score (main ranking criterion), projection quality safeguards (trustworthiness/continuity $\geq 0.80$), and noise ratio analysis for density-based methodologies.  Stability validation employs bootstrap resampling (5 iterations) \cite{maqbool2006stability} to calculate the Adjusted Rand Index (ARI) and Normalized Mutual Information (NMI) throughout several executions.  This comprehensive strategy emphasizes thorough partitions rather than statistical purity, all while upholding stringent quality standards.

\begin{table}[!htbp]
\vspace{-5pt}
\centering
\caption{Complete sweep configuration.}
\label{tab:method_config}
\small
\begin{tabularx}{\linewidth}{l X}
\toprule
\textbf{Component} & \textbf{Specification} \\
\midrule
Embedding model & BGE-large-en-v1.5 + GTE-large-en-v1.5 (dual, concatenated, 2048-dim), codebook $K_c=47$ (auto-selected) \\
Feature types & TF-IDF weighted codebook counts (L2 normalized), quantized embeddings, SVD variants \\
Projection spaces & RAW, SVD, UMAP (4/8/16-D, varying hyperparameters) \\
Clustering algorithms & K-means, Agglomerative (multiple linkages), DBSCAN, OPTICS \\
Selection criteria & Silhouette-based selection with trustworthiness/continuity guardrails and stability validation, with complete assignment methods prioritized for final selection \\
\bottomrule
\end{tabularx}
\vspace{-10pt}
\end{table}

\subsection{Codebook Diagnostics}
I automatically scan codebook size $K_c$ across candidates $\{32, 48, ..., 1024\} \cup \{\sqrt{n}, 1.5\sqrt{n}, 2\sqrt{n}\}$ using the adjusted silhouette score $S_{adj} = S - 0.6 \cdot r_{singleton} - 0.8 \cdot r_{empty} - 0.2 \cdot \text{Gini}$, where $r_{singleton}$ and $r_{empty}$ denote singleton and empty cluster ratios. Four clustering algorithms (K-means variants, Agglomerative with Ward linkage) are evaluated on PCA-whitened dual model embeddings (2048-dim reduced to 95\% variance, typically ~1800 dimensions). The selected configuration (MiniBatch K-means with batch\_size=1024, $K_c=47$) demonstrates balanced cluster utilization with low singleton occurrence and semantic coherence across axes. Representative semantic clusters include \{\textit{biofabrication}, \textit{tissue-engineering}, \textit{living-materials}\} and \{\textit{posthumanism}, \textit{new-materialism}, \textit{relational-ontology}\}.

\subsection{Technical Implementation}
\textbf{Reproducibility Framework.} All stochastic components use fixed random states, ensuring deterministic results. The four-stage pipeline processes (1) dual token embeddings (BGE-large-en-v1.5 + GTE-large-en-v1.5 concatenated, 2048-dim), (2) PCA-preprocessed K-means codebook construction ($K_c=47$, auto-selected), (3) TF-IDF weighted representations with L2 normalization, and (4) systematic evaluation across 800+ algorithm-space combinations within computational limits. Key configuration parameters include PHASEC\_MAX\_TRIALS=800, PHASEC\_K\_LIST=[2,3,...,15], PHASEC\_BOOTSTRAP\_REPS=5, and random seeds (Python=42, NumPy=42, sklearn models=42). UMAP projections use cosine metric for TF-IDF features with hyperparameter grids of neighbors $\{10,15,30\}$, distances $\{0.01,0.1,0.5\}$, and dimensions $\{4,8,16\}$.

\section{Results}

I systematically assess $K\in[2,15]$ informed by the cluster number selection literature \cite{halkidi2001clustering,milligan1985examination,kaufman1990finding}.  The top limit $K \le 15$ reflects three considerations. First, \textbf{statistical validity} requires adequately populated clusters for dependable internal metrics \cite{halkidi2001clustering}, and as a rule of thumb we allow 4-5 samples per cluster, giving $K \le N/4 \approx 20$ for our corpus of $N=81$. Second, \textbf{interpretive manageability} favors fewer clusters, since partitions beyond $K=15$ raise the cognitive load of categorical interpretation in an atlas setting. Third, \textbf{dimensional constraints} relative to the 13 analytic axes mean that an excessive $K$ can erode distance-based separation.  This principled range mitigates issues related to arbitrary constraint selection while preserving atlas usability.

The optimal configuration is Agglomerative (average) at $k=15$ on 4D UMAP, achieving silhouette $0.664 \pm 0.008$ (5 seeds) with high neighborhood preservation (trustworthiness/continuity $\approx 0.81 \pm 0.01$). My $K=15$ selection aligns with theoretical guidelines, since each cluster contains 5.4 samples on average (above the 4-5 minimum) while remaining interpretively manageable. Alternative criteria (Gap statistic, Elbow method) suggest optimal ranges of $K \in [12,18]$ and $K \in [8,14]$ respectively, with convergence around our selected value validating the methodological approach. KMeans peaks near $k\in\{14,15\}$ but remains below hierarchical performance. Density methods achieve higher silhouettes primarily via noise exclusion, which I report but do not use for exhaustive atlas labeling. Bootstrap resampling confirms result stability (coefficient of variation <2\%).

\subsection{Clustering and Visualization}
I report the clustering sweep exactly as run on the released features and scripts. Features are TF-IDF weighted codebook counts with row-wise L2 normalization, produced by our systematic preprocessing pipeline. My systematic sweep evaluates \emph{projection spaces} (RAW/SVD/UMAP with varying dimensionality and neighborhood parameters) and algorithms (KMeans, Agglomerative, DBSCAN, OPTICS). Metric conventions, primary ranking, and handling of density-method noise follow §\ref{subsec:conv}.

\paragraph{Best (internal validity).}
The optimal configuration by \emph{multi-metric evaluation} is \emph{Agglomerative (average linkage)} with $k{=}15$ on 4D UMAP (10 nearest neighbors, minimum distance 0.01). The silhouette is $\mathbf{0.664 \pm 0.008}$, satisfying the trustworthiness/continuity guardrails ($\approx 0.81$).

\begin{table}[!htbp]
\vspace{-5pt}
\centering
\caption{Best runs per algorithm using multi-metric evaluation (silhouette, trustworthiness/continuity). Density methods report noise ratio but are excluded from exhaustive atlas labeling.}
\label{tab:algo_bench}
\begingroup
\scriptsize
\setlength{\tabcolsep}{6pt}
\renewcommand{\arraystretch}{1.1}
\begin{tabular}{l c c c c l}
\toprule
Algorithm & \#Clusters & Noise (\%) & Silhouette & Trust./Cont. & UMAP Configuration \\
\midrule
K-means        & 15 & 0.0  & 0.483 & 0.805/0.812 & 4D (neighbors=10, min\_dist=0.01) \\
Agglomerative  & 15 & 0.0  & \textbf{0.664} & 0.805/0.812 & 4D (neighbors=10, min\_dist=0.01) \\
DBSCAN         &  2 & 71.6 & 0.887 & 0.795/0.833 & 8D (neighbors=10, min\_dist=0.1) \\
OPTICS         &  5 & 59.3 & 0.809 & 0.801/0.814 & 4D (neighbors=15, min\_dist=0.1) \\
\bottomrule
\end{tabular}
\endgroup
\vspace{-7pt}
\end{table}

\paragraph{Viewer labeling (communication-first).}
For public exploration, I also provide a lower-$k$ labeling to keep the map readable. This is a communication-oriented choice distinct from the silhouette-maximizing run. I avoid mixing the two objectives.

\begin{figure*}[!htbp]
\vspace{-10pt}
  \centering
  \includegraphics[width=\textwidth]{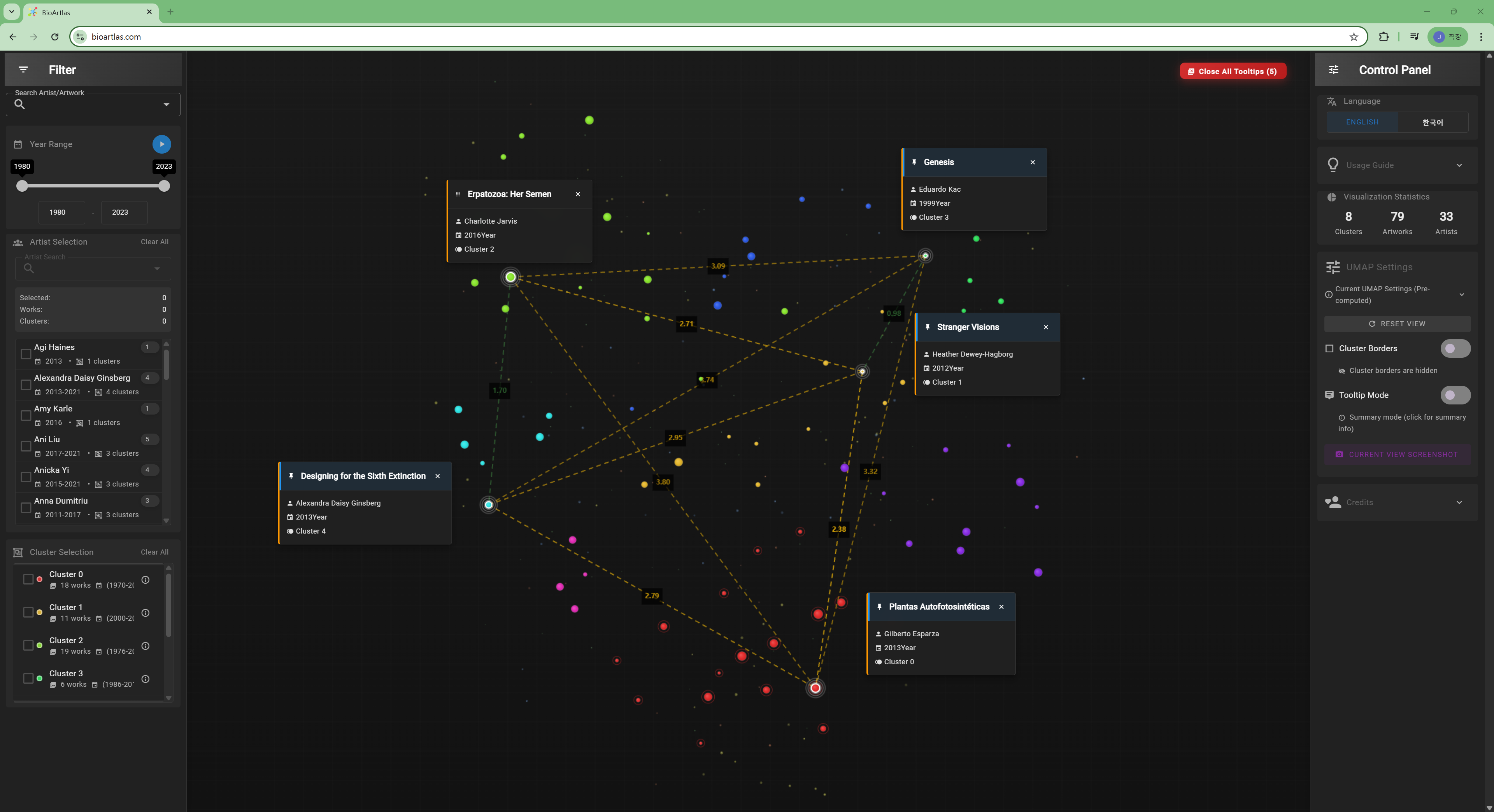}
  \caption{BioArtlas interactive visualization interface. \url{https://www.bioartlas.com}}
  \label{fig:teaser}
  \vspace{-15pt}
\end{figure*}

\subsection{Cluster Analysis and Discovered Patterns}

Pattern discovery integrates quantitative clustering outcomes with qualitative visual analysis using an interactive web interface (Figure~\ref{fig:teaser}).  The viewer facilitates systematic examination of cluster borders, artist trajectories, and temporal relationships, bolstering the interpretations below while preserving analytical objectivity via k-NN membership and rank-based proximity metrics.

\textbf{Methodological Cohesion: Stelarc's Body Intervention Art.} Cluster 4 shows strong artist-specific cohesion. Stelarc's four works share consistently high within-cluster proximity in terms of mutual $k$-NN membership and small rank displacement, rather than relying on absolute map distances. \textit{Suspensions} (1976), \textit{Third Hand} (1980), \textit{Stomach Sculpture} (1993), and \textit{Ear on Arm} (2006) jointly indicate a persistent methodological domain across three decades, centered on cyborg embodiment and posthuman performance.

\textbf{Segmented Distribution of Tissue Culture Art.} Tissue-culture works appear in two nearby regions with substantial neighborhood overlap, indicating related but distinguishable foci. Cluster 1 includes Marta de Menezes's early works and Špela Petrič's plant-human studies. Cluster 13 encompasses Catts \& Zurr's mature works alongside recent DNA-based projects.

\textbf{Multi-cluster Distribution of Individual Artists.} Several artists' works occupy distinct local neighborhoods with low mutual $k$-NN overlap across clusters, revealing methodological evolution over time. Joe Davis's four works and Eduardo Kac's three works each form separate neighborhood sets whose nearest-neighbor composition shifts across periods, indicating technical and conceptual diversification within the same practice without relying on absolute map distances.

\textbf{Trans-temporal Conceptual Affinity.} Clustering prioritizes conceptual similarity over chronology. Joe Davis's \textit{Poetica Vaginal} (1986) and Jenna Sutela's \textit{nimiia cétiï} (2018) show high reciprocal $k$-NN membership and small rank-displacement, indicating a shared microbial-linguistic focus despite a 32-year gap. I report this affinity via neighborhood overlap and rank structure rather than absolute UMAP distances.

\section{Conclusion}

My axis-aware methodology structures the multidimensional complexity of bioart into an interpretable map, attaining a silhouette of $0.664 \pm 0.008$ across a systematic assessment of over 800 configurations.

\textbf{Key Contributions:} (1) \textbf{axis-aware representation learning} preserving semantic distinctions across thirteen heterogeneous dimensions while enabling cross-dimensional comparison, (2) \textbf{domain-informed codebook construction} grouping related concepts into unified clusters and addressing cultural terminology polysemy, (3) \textbf{systematic evaluation framework} explicitly separating analytical optimization from communicative design, and (4) \textbf{discovery of four organizational patterns} (methodological cohesion, technique segmentation, artistic evolution, and trans-temporal affinities) that complement art-historical analyses.

\textbf{Broader Impact:} This approach offers a model for computational cultural study in areas characterized by multidimensional complexity.  The modular design facilitates systematic expansion across geographic borders, multi-annotator validation, and cross-domain extension.

\textbf{Future Works:} Present limitations include Western-centric bias and single-annotator labeling.  I intend to enlist specialists and curators for multi-annotator validation of the 13-dimensional annotations, enabling inter-rater reliability assessment.  Geographic expansion will reach bioart communities in the Asia-Pacific, Latin America, and Africa to reduce bias while integrating contemporary AI/ML works.

\bibliographystyle{unsrtnat}
\bibliography{references}

\end{document}